\documentclass{aa}  

\usepackage{graphicx, hyperref}
\hypersetup{colorlinks = true,
            citecolor = {blue},
            }

\usepackage{txfonts}
\newcommand{\hi}{H\,{\textsc{i}}}
\newcommand{\hii}{H\,{\textsc{ii}}}
\newcommand{\nii}{N\,{\textsc{ii}}}
\newcommand{\oiii}{O\,{\textsc{iii}}}
\newcommand{\oii}{O\,{\textsc{ii}}}
\newcommand{\neiii}{Ne\,{\textsc{iii}}}

\newcommand{\jwst}{{\em JWST}}
\newcommand{\alma}{{\em ALMA}}

\begin{document}

   \title{Novel $z\sim~10$ auroral line measurements extend the gradual offset of the FMR deep into the first Gyr of cosmic time}
    \titlerunning{Rest-frame optical lines and metallicities for galaxies at $z\approx 10$}
    \authorrunning{Pollock et al.}

   %\subtitle{I. Overviewing the $\kappa$-mechanism}

   \author{Clara~L.~Pollock
          \inst{1,2}
          \and
          Rashmi~Gottumukkala\inst{1,2}
          \and
          Kasper~E.~Heintz \inst{1,2,3}
          \and
          Gabriel~B.~Brammer\inst{1,2}
          \and
          Guido~Roberts-Borsani\inst{4}
          \and
          Pascal~A.~Oesch\inst{3,1,2}
          \and
          Joris~Witstok\inst{1,2}
          \and
          Karla~Z.~Arellano-C\'{o}rdova\inst{5}
          \and
          Fergus~Cullen\inst{5}
          \and
          Dirk~Scholte\inst{5}
          \and
          Chamilla~Terp\inst{1,2}
          \and
          Lucie~Rowland\inst{6}
          \and 
          Albert~Sneppen\inst{1,2}
          \and
          Kei~Ito\inst{1,7}
          \and 
          Francesco~Valentino\inst{1,7}
          \and
          Jorryt~Matthee\inst{8}
          \and
          Darach~Watson\inst{1,2}
          \and
          Sune~Toft\inst{1,2}
          }

   \institute{Cosmic Dawn Center (DAWN), Denmark
        \and
        Niels Bohr Institute, University of Copenhagen, Jagtvej 128, 2200 Copenhagen N, Denmark
        \and
        Department of Astronomy, University of Geneva, Chemin Pegasi 51, 1290 Versoix, Switzerland
        \and
        Department of Physics \& Astronomy, University College London, London, WC1E 6BT, UK
        \and
        Institute for Astronomy, University of Edinburgh, Royal Observatory, Edinburgh, EH9 3HJ, UK
        \and
        Leiden Observatory, Leiden University, P.O. Box 9513, NL-2300 RA Leiden, the Netherlands
        \and
        DTU Space, Technical University of Denmark, Elektrovej 327, DK2800 Kgs. Lyngby, Denmark
        \and
        Institute of Science and Technology Austria (ISTA), Am Campus 1, 3400 Klosterneuburg, Austria}

   \date{June 2025}

% \abstract{}{}{}{}{} 
% 5 {} token are mandatory
 
  \abstract
   {The mass assembly and chemical enrichment of the first galaxies provide key insights into their star-formation histories and the earliest stellar populations at cosmic dawn. Here we compile and utilize new, high-quality spectroscopic \jwst/NIRSpec Prism observations from the \jwst\ archive. In particular, we extend the wavelength coverage beyond the standard pipeline cutoff ($5.3\mu$m) up to 5.5\,$\mu$m, which enable for the first time a detailed examination of the rest-frame optical emission-line properties for galaxies at $z\approx 10$. Crucially, the improved calibration allows us to detect H$\beta$ and the [\oiii]\,$\lambda\lambda 4959,5007$ doublet and resolve the auroral [O\,{\sc iii}]\,$\lambda 4363$ line for the 11 galaxies in our sample ($z=9.3-10.0$) to obtain direct $T_e$-based metallicity measurements. We find that the interstellar medium (ISM) of all galaxies show high ionisation fields and electron temperatures, with derived metallicities in the range $12+\log {\rm (O/H)} = 7.1 - 8.3$ (3--50\% solar), consistent with previous strong-line diagnostics based on \jwst\ data at high redshifts. We derive an empirical relation for $M_{\rm UV}$ and 12+log(O/H) at $z\approx 10$, useful for future higher-redshift studies, and show that the sample galaxies are `typical' star-forming galaxies though with relatively high specific star-formation rates (median sSFR = SFR$_{\rm H\beta}$/$M_\star$ = 58\,Gyr$^{-1}$) and with evidence for bursty star formation on 10\,Myr vs 100\,Myr timescales ($\log_{10}({\rm SFR_{10}/SFR_{100}}) = 0.4$). Combining the rest-frame optical line analysis and detailed UV to optical spectro-photometric modelling, we determine the mass-metallicity relation (MZR) and the fundamental-metallicity relation (FMR) of the sample, pushing the previous redshift frontier of these measurements to $z=10$. These results, together with literature measurements, point to a gradually decreasing MZR at higher redshifts, with a break in the FMR at $z\approx 3$, decreasing to metallicities $\approx 3\times$ lower at $z=10$ than observed in galaxies during the majority of cosmic time at $z=0-3$, likely caused by massive pristine gas inflows diluting the observed metal abundances during early galaxy assembly at cosmic dawn.
   }
   %We use these metallicity measurements together with estimates of the star-formation rates from H$\beta$ and stellar masses from rest-frame UV to optical spectro-photometric modelling to determine the fundamental metallicity scaling relations of galaxies, 
   %pushing the previous redshift frontier of these measurements to $z=10$. 
   % This points to X and Y with implications for Z.}

   \keywords{high-redshift galaxies --
                galaxy formation, evolution 
               }

   \maketitle
%
%--------------------------------------------------------------

\section{Introduction}

% Broad intro to galaxy assembly and chemical evolution
Probing the chemical enrichment of the first galaxies in the early Universe is key to constrain the physical processes that regulate their formation and evolution via gas inflows and outflows, star formation, and the synthesis of heavier elements in stellar cores. This process has been characterized by the scaling relation between the stellar mass and chemical abundance of oxygen, known as the mass-metallicity relation \citep{Lequeux79,Tremonti04,Lee06,Erb06,Kewley08}. Perhaps more fundamental is the additional scaling with the star-formation rate, called the fundamental-metallicity relation \citep{Ellison08,Mannucci10,Lilly13,Andrews13,Curti20}, which has been found to be constant out to $z\approx 3$ \citep{Sanders21}, throughout the majority of cosmic time. This implies universal galaxy growth, potentially representing the universal scaling between the molecular gas mass surface density and star formation \citep{Baker23}.

% How rest-frame optical lines can help us constrain metallicities
Rest-frame optical spectroscopy of strong nebular emission lines enables key insights into the physical properties and chemical enrichment of star-forming regions in galaxies through cosmic time \citep[see e.g.,][for reviews]{Maiolino19,Kewley19}. 
With the advent of \jwst\ and its suite of sensitive, near-infrared spectroscopic capabilities \citep{Jakobsen22}, it is now possible to measure and constrain rest-frame optical nebular emission lines out to very high redshifts. The first studies of a statistical sample of galaxies at $z\approx 8$ with detected rest-frame optical emission lines \citep{Langeroodi23,Heintz23_FMR, Sanders24} revealed a significantly decreasing metallicity at a given stellar mass compared to past measurements out to $z\approx 3$. This was later confirmed by larger \jwst\ samples \citep{Nakajima23,Curti24,Morishita24}, and for very low stellar mass ($M_\star \sim 10^{6}\,M_\odot$) \citep{Chemerynska24b} to near-solar metallicities \citep{Rowland25}. Perhaps most surprising was the apparent offset from the fundamental-metallicity relation towards 0.5 dex lower metallicities for a given stellar mass and star-formation rate \citep{Heintz23_FMR}, hinting at a transition period for high-redshift galaxies dominated by pristine gas inflows, though the exact redshift is still debated \citep[constrained to within $z\sim 4-8$;][]{Nakajima23,Curti24}. Strong-line measurements further enable the determination of the relative elemental abundances and the ionisation fields of the interstellar medium (ISM) \citep[e.g.,][]{Schaerer22,ArellanoCordova22,ArellanoCordova24,Heintz23_JWSTALMA,Trump23,Shapley23b} for some of the most distant galaxies known out to $z\approx 9$. 

Most of the previous metallicity measurements of high-redshift galaxies have relied on strong-line diagnostics due to the inaccessibility of the temperature-sensitive [\oiii]\,$\lambda 4363$ auroral lines, either due to its intrinsic faintness or because it is blended with H$\gamma$ in low-resolution spectra. 
%Several methods can be used to calculate the gas-phase metallicity, including the direct `$T_e$-method' which uses the temperature-sensitive auroral line [\oiii] $\lambda 4363$. 
However, a growing number of direct measurements using the direct `$T_e$-method' are now being reported based on \jwst\ grating spectra \citep{Laseter2024, Sanders24, Scholte25, Cullen25}, which has been crucial to calibrate the empirical strong-line relations for the majority of galaxies with no auroral line detection. 
%In the standard NIRSpec Prism wavelength coverage, necessary strong lines ([\oiii], H$\beta$) are not available at $z\gtrsim9.5$; making the direct method impossible, and leaving few available calibrations, e.g. high-scatter relation Ne3O2 based on [\neiii] and [\oii], which are not guaranteed to be measured at such high-z, and may trace ionisation more than oxygen abundance. 
%additional dependencies on ionisation/nitrogen abundance? 
Despite the growing number of $T_e$-based 12+log(O/H) estimates, there are still relatively few at $z > 9$. Here we utilize new, custom reductions of recent \jwst/NIRSpec Prism spectroscopic data, expanding the wavelength coverage out to $5.5\mu$m (up from $5.3\mu$m), to investigate the rest-frame UV and optical lines up to and including the [O\,{\sc iii}]\,$\lambda \lambda 4959,5007$ line doublet at $z=10.0$. These we use to benchmark against existing high-redshift strong-line diagnostics and determine fundamental galaxy scaling relations up to the highest redshifts achievable by \jwst/NIRSpec. 

We structure the paper as follows. In Sect.~\ref{sec:obs} we present our new spectroscopic reductions of archival \jwst\ data and detail the sample compilation. In Sect.~\ref{sec:res} we describe the emission-line analyses and spectro-photometric modelling and in Sect.~\ref{sec:gal} we present the new empirical galaxy scaling relations at $z\approx 10$. In Sect.~\ref{sec:disc} and Sect.~\ref{sec:conc} we discuss and conclude on our work. Throughout the paper we assume the standard concordance cosmology, with a flat, $\Lambda$CDM-dominated Universe. We adopt the cosmological parameters from \citet{Planck18} and the solar abundances from \citet{Asplund09} with $12+\log({\rm O/H})_\odot = 8.69$. 
%and a \citet{Chabrier03} initial mass function (IMF). 

% list/define all line ratios used? 
%--------------------------------------------------------------------
\section{Observations} \label{sec:obs}
%--------------------------------------------------------------------

% Detail DJA here, new v4 reductions compared to v2 (Heintz+) and v3 (de Graaff+). 

For this work, we compile and utilize the archival \jwst\ data reduced and processed as part of the DAWN \jwst\ Archive (DJA)\footnote{\url{https://dawn-cph.github.io/dja/}}. This online repository contains reduced images, photometric catalogs, and spectroscopic data for public \jwst\ data products. The raw spectroscopic data are retrieved from MAST, before they are processed with {\tt MSAExp} \citep{Brammer_msaexp}\footnote{\url{github.com/gbrammer/msaexp}}, as detailed in \citet[][for version 2, {\tt v2}]{Heintz25} and \citet[][for {\tt v3}]{DeGraaff24}, respectively.

Here we present and utilize the most recent spectra from {\tt v4} of the DJA spectroscopic archive \citep[see also][]{Valentino25}. We focus on the spectra observed with the \jwst/NIRSpec Prism configuration \citep{Jakobsen22}, which nominally covers $\lambda = 0.6-5.3\,\mu$m with a resolving power $\mathcal{R} \approx 100$. 
Briefly, DJA-{\tt v4} uses updated \jwst\ reference files, and includes a bar shadow correction, overall improving the absolute and colour-dependent flux calibration. Newly, we extend the {\tt wavelengthrange} reference file\footnote{\url{https://jwst-pipeline.readthedocs.io/en/latest/jwst/references_general/wavelengthrange_reffile.html}} to include the 2nd and 3rd order Prism spectra, which were cut from the \jwst\ pipeline to avoid the higher-order contamination on the primary first spectral order \citep{Jakobsen22}. These, however, appear at predictable locations and intensities, and are therefore now carefully calibrated into the DJA-{\tt v4} reductions to include the full response of the telescope and thereby recover the full spectrum recorded on the detector \citep[see also][for details on the same procedure on the higher-resolution grating spectra]{Valentino25}. This effectively extends the wavelength coverage of NIRSpec Prism to $5.5\mu$m. The final 1D spectra were optimally extracted \citep{Horne86} and the flux calibrations are generally accurate to the photometry (see below). The {\tt v4} reductions have been run and processed on the entire DJA, including also more recent public \jwst/NIRSpec observations, and can be found on the dedicated online repository\footnote{\url{https://s3.amazonaws.com/msaexp-nirspec/extractions/public_prelim_v4.2.html}}. %% Check when/if we want to go public with this

Consequently, we can now detect and constrain the rest-frame optical emission lines such as the crucial [\oiii]\,$\lambda\lambda 4959,5007$ doublet up $z=10.0$ with \jwst/NIRSpec  (instead of only up to $z\approx9.5$ with standard pipeline products). Due to increasing spectral resolution with wavelength, and the generally higher delivered resolving power, we measure up to $\mathcal{R} \approx 500$ at $5.5\mu$m. This increased resolution is sufficient to spectroscopically resolve the [\oiii]\,$\lambda 4363$ auroral line transition from H$\gamma$ at $z\gtrsim 9.0$, allowing direct, $T_e$-based metallicity estimates at $z=9.0-10.0$. We compile a set of 11 galaxies from DJA-{\tt v4} with \jwst/NIRSpec Prism spectra, by requiring a robust detection of H$\beta$ (${\rm S/N} > 5$) and wavelength coverage of the [\oiii]\,$\lambda\lambda 4959,5007$ doublet. This mainly includes galaxies observed as part of the major \jwst\ spectroscopic surveys: CAPERS \citep[prog. ID: 6368, PI: Dickinson;][]{Kokorev25}, JADES \citep[prog. IDs: 1181 and 3215, PI: Eisenstein;][]{Bunker23_jades,Eisenstein23a, Eisenstein23b}, UNCOVER \citep[prog. ID: 2561, PIs: Labbe and Bezanson;][]{Bezanson23}, RUBIES \citep[prog. ID: 4233, PIs: de Graaff \& Brammer;][]{DeGraaff24} and RXJ DDT (prog. ID: 2767, PI: Kelly). %CEERS \citep[prog. ID: 1345, PI: Finkelstein;][]{Finkelstein23}, 
The final sample is summarized in Table~\ref{tab:lines}. 

\begin{table*}
    %\begin{center}
     \renewcommand{\arraystretch}{1.0} % Default value: 1
      \setlength{\tabcolsep}{5pt} % Default value: 6pt
            \caption{Sample overview with derived redshifts and rest-frame optical emission-line fluxes, or 3$\sigma$ upper limits.}
            % \resizebox{\paperwidth}{!}{%
        \begin{tabular}{ l c c c c c c c } 
         \hline \hline    
            Source & $z_{\rm spec}$ & [\oii]\,$\lambda\lambda 3726,29$ & [\neiii]\,$3869$ & H$\gamma$ & [\oiii]\,$\lambda 4363$ & H$\beta$ & [\oiii]\,$\lambda 5007$ \\
            \hline   
            CAPERS-EGS-25297 & $9.9381 \pm 0.0003$ & $73.0^{+3.3}_{3.3}$ & $71.8^{+3.1}_{-3.1}$ & $75.7^{+4.1}_{-4.0}$ & $28.2^{+3.8}_{-3.8}$ & $207.2^{+6.9}_{-6.8}$ & $1172.6^{+9.5}_{-9.7}$ \\
            
            UNCOVER-2561-13151 $^{(a*)}$  & $9.8026 \pm 0.0003$ & $7.0^{+2.2}_{-2.2}$ & $22.6^{2.3}_{-2.3}$ & $15.5^{+2.5}_{-2.6}$ & $<7.7$ & $52.7^{+3.6}_{-3.5}$ & $288.4^{+7.9}_{-7.6}$ \\
            
            JADES-GN-55757 &  $9.7498 \pm 0.0006$ & $16.7^{+4.4}_{-4.5}$ &  $18.1^{+4.9}_{4.9}$ & $47.9^{+5.3}_{-5.3}$ & $<15.8$ & $72.2^{+6.9}_{-6.9}$ & $339.4^{+12.0}_{-11.6}$ \\
            
            UNCOVER-2561-22223 $^{(b*)}$  & $9.5704 \pm 0.0008$  & $< 10.04$ & $10.3^{+3.4}_{-3.3}$& $22.9^{+4.1}_{-4.0}$ & $<12.6$& $74.7^{+5.7}_{-5.9}$ & $229.7^{+7.5}_{-7.6}$\\   
            
            RXJ2129-2767-11027 $^{(c*)}$  & $9.5127 \pm 0.0001$ & $1.9^{+0.4}_{-0.4}$ & $2.1^{+0.4}_{-0.4}$ & $4.2^{+0.5}_{-0.4}$ & $1.9^{+0.5}_{-0.5}$  &$7.3^{+0.6}_{-0.6}$ & $40.5^{+0.8}_{-0.7}$ \\
            
            JADES-GS-265801 $^{(d)}$  & $9.4437 \pm 0.0001$ & $9.4^{+0.9}_{-0.8}$ & $27.0^{1.0}_{-0.9}$ & $34.5^{+1.0}_{-1.0}$& $14.1^{+1.0}_{-1.0}$ & $81.5^{+1.3}_{-1.3}$ & $480.3^{+2.0}_{-2.0}$\\   
            
            CAPERS-EGS-87132 & $9.3833 \pm 0.0002$ & $17.6^{+2.9}_{-2.9}$ & $25.3^{+3.1}_{-3.1}$& $20.6^{+3.6}_{-3.6}$ & $12.3^{+3.4}_{-3.4}$  & $38.4^{+4.0}_{-4.1}$ & $282.8^{+5.7}_{-5.5}$ \\
            
            JADES-GN-3990 & $9.3812 \pm 0.0002$ & $16.6^{+2.6}_{-2.5}$ & $22.2^{+2.5}_{-2.5}$ & $25.4^{+3.0}_{-3.0}$ & $12.3^{+2.7}_{-2.9}$& $54.6^{+3.6}_{-3.6}$ & $366.4^{+4.7}_{-4.8}$ \\   
            
            UNCOVER-2561-3686 $^{(e*)}$  & $9.3202 \pm 0.0004$ & $73.6^{+10.5}_{-10.5}$ &  $79.1^{+10.7}_{-10.5}$ & $48.4^{+10.4}_{-10.5}$ & $<32.9$ & $49.2^{+12.5}_{-12.5}$ & $639.5^{+16.0}_{-15.5}$ \\
            
            RUBIES-UDS-833482 & $9.3042 \pm 0.0003$ & $93.1^{+13.9}_{-13.9}$ & $70.3^{+12.7}_{-12.9}$ & $91.9^{+15.9}_{-15.3}$ & 
            $<36.2$ & $165.7^{+16.0}_{-15.7}$ & $1595.2^{+26.2}_{-26.2}$ \\   
            
            CAPERS-UDS-22431 & $9.2717 \pm 0.0001$ & $27.1^{+3.1}_{-3.2}$ & $44.5^{+3.4}_{-3.5}$ & $35.8^{+3.7}_{-3.8}$ & $19.8^{+3.7}_{-3.7}$ & $88.2^{+4.7}_{-4.7}$ & $727.8^{+6.9}_{-6.9}$ \\
            \hline \hline  
        \end{tabular}  \\\\
       %\end{center}
       % }
    \textbf{Notes.} Selected galaxies have reported/analysed in previous works: $^{(a)}$ \cite{Zitrin14, RobertsBorsani23, Fujimoto23_uncover} $^{(b)}$ \cite{Fujimoto23_uncover} $^{(c)}$ \cite{Williams23} $^{(d)}$ \cite{Curti25} $^{(e)}$ \cite{Boyett24, Atek23b, Castellano23, Fujimoto23_uncover}. $^{(*)}$ Lensed galaxies, the reported line fluxes are not corrected for magnification.\\ The [\oii]\,$\lambda\lambda 3726,3729$ line doublet is blended so we report the total line fluxes. We use the theoretical ratio of [\oiii]\,$\lambda 5007$/[\oiii]\,$\lambda 4959 \sim 2.98$ to fix the line flux of [\oiii]\,$\lambda4959$. All line fluxes are uncorrected for dust attenuation and in units of $10^{-20}$\,erg\,s$^{-1}$\,cm$^{-2}$. 
        \label{tab:lines}
\end{table*}

%For each galaxy, we also include photometric data obtained primarily with \jwst/NIRCam through DJA \citep{Valentino23}. This we use to photometrically calibrate the flux density of the spectra and
%to estimate the physical sizes, characterized as the effective half-light radius in the F150W band (approx. rest-frame $1500\,\AA$ at $z\approx 9-10$), and 
%for the spectro-photometric modelling of the spectral energy distribution (SED) of each source.

For each galaxy, the photometric data obtained primarily with \jwst/NIRCam through DJA \citep{Valentino23} was inspected, in order to photometrically calibrate the flux density of the spectra. Of the 11 galaxies in the sample, only two require a positive (>1) scaling, consistent with slit loss. For these objects the spectra were calibrated using {\sc Bagpipes} during the spectro-photometric modelling of the spectral energy distribution (SED), and the calibrated spectra were used for line-fitting. For the remainder of the sample, the photometry is either consistent with spectroscopy; not requiring rescaling, or lower than the spectrum for no obvious physical reason from inspecting the morphology. The analysis for these galaxies is carried out directly on the DJA-reduced spectra.

% Describe Galaxy compilation + add table.

\section{Analysis and results} \label{sec:res}

\subsection{Emission-line modelling}

% Present line flux measurements Fig. of CAPERS example? + table + Fig. showing line ratios (a la Schaerer+). 

We model the most prominent rest-frame optical nebular emission lines detected in the spectra, from the (unresolved) [O\,{\sc ii}]\,$\lambda\lambda 3726,3729$ doublet to the [O\,{\sc iii}]\,$\lambda\lambda 4959,5007$ doublet line transitions. Due to the expanded wavelength coverage, the latter can now be measured up to $z=10.0$, while simultaneously resolving the auroral [O\,{\sc iii}]\,$\lambda 4363$ line from H$\gamma$. We determine the line fluxes or upper bounds on the emission lines by modelling the full set of lines jointly with Gaussian line profiles (at vacuum wavelengths), tying the redshift and intrinsic width (convolved with the Prism spectral resolution). We model the underlying continuum with a simple first-order polynomial. To estimate the model parameters, we use Dynamic Nested Sampling from {\sc dynesty} \citep{Speagle20}, which allows for more complex distributions to be modelled \citep{Koposov2022_zenodo, Higson2019}.

%To estimate the model parameters, we use Dynamic Nested Sampling from {\sc dynesty} \citep{Speagle20}, which simultaneously models the evidence and posterior distribution by integrating over the prior in `shells' of constant likelihood \citep{Koposov2022_zenodo}. Nested sampling typically allows for more complex distributions to be modelled than MCMC methods, and can provide more accurate estimates on both the statistical and sampling uncertainties associated with each run. Throughout the course of the fit, the number of `live points' is allowed to vary, resulting in a more efficient and accurate sampling of the parameter space \citep{Higson2019}. %PRIORS

\begin{figure*}
    \centering
    \includegraphics[width=1.0\textwidth]{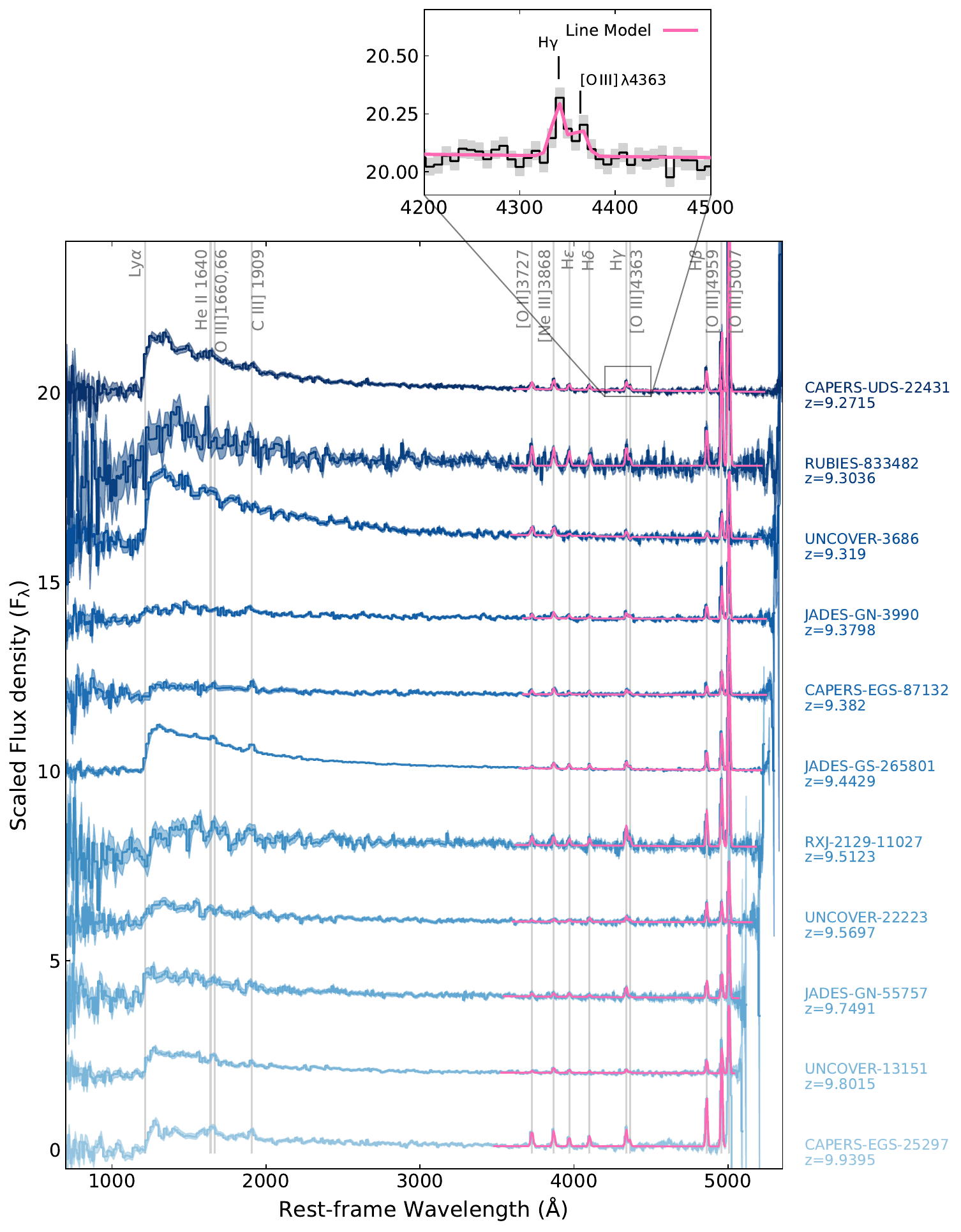}
    \caption{\jwst/NIRSpec Prism 1D spectroscopy of the sample, shown in shades of blue, with the associated error spectrum as a shaded region. The position of the main line transitions are highlighted by grey lines. The inset zooms in on the spectroscopically resolved H$\gamma$+[\oiii]\,$\lambda 4363$ emission lines, with the best-fit continuum and emission-line model for each galaxy shown by the pink lines.}
    \label{fig:allspec}
\end{figure*}

An example of the line fitting, highlighting the spectroscopically resolved [O\,{\sc iii}]\,$\lambda 4363$ line, is shown in Fig.~\ref{fig:allspec}. We find that the observed spectral resolution is typically 1.20-1.70 higher than the nominal \jwst/NIRSpec Prism curve at any given wavelength \citep[see also e.g.,][]{DeGraaff24}. The derived line fluxes and derived redshifts are summarized in Table~\ref{tab:lines}.\\
%and the line fitting for all galaxies are shown in the Appendix.  
The observed line fluxes were corrected for dust-reddening using SED-derived extinction $\rm A_V$ values, from the {\sc Bagpipes} fits outlined in Sect.~\ref{sec:bagpipes}. Emission line fluxes are corrected assuming a modified \cite{Calzetti2000} dust curve (see \cite{Salim18,Salim20}), with slope $\delta=-0.3$ and $R_V=3.0$. We find the sample is generally dust-poor, with a maximum $A_V = 0.36$, and mean $A_V \sim 0.2$. The dust-corrected line fluxes are used for emission line ratios, as well as for derivations of $T_e$, $\rm 12+log(O/H)$, and $\rm SFR(H\beta)$.

%\begin{figure*}
%    \centering
%    \includegraphics[width=0.8\textwidth]{Figs/Example_spectrum.png}
%    \caption{JWST/NIRSpec Prism 1D example spectroscopy of CAPERS-EGS-25297 at $z=9.94$. The 2D spectrum is shown on top, the 1D spectrum is shown in the centre as the black curve, with the associated error spectrum in grey. The newly extended region of the spectrum is shown in blue. The inset zooms in on the spectroscopically resolved H$\gamma$+[\oiii]\,$\lambda 4363$ line transitions, with the best-fit continuum and emission-line model shown by the pink line. The \jwst/NIRCam photometric data points are overplotted as the blue circles. 
%    The bottom panel shows the associated filter transmission curves.}
%    \label{fig:linefitex}
%\end{figure*}

\begin{figure*}
    \centering
    \includegraphics[width=1.0\textwidth]{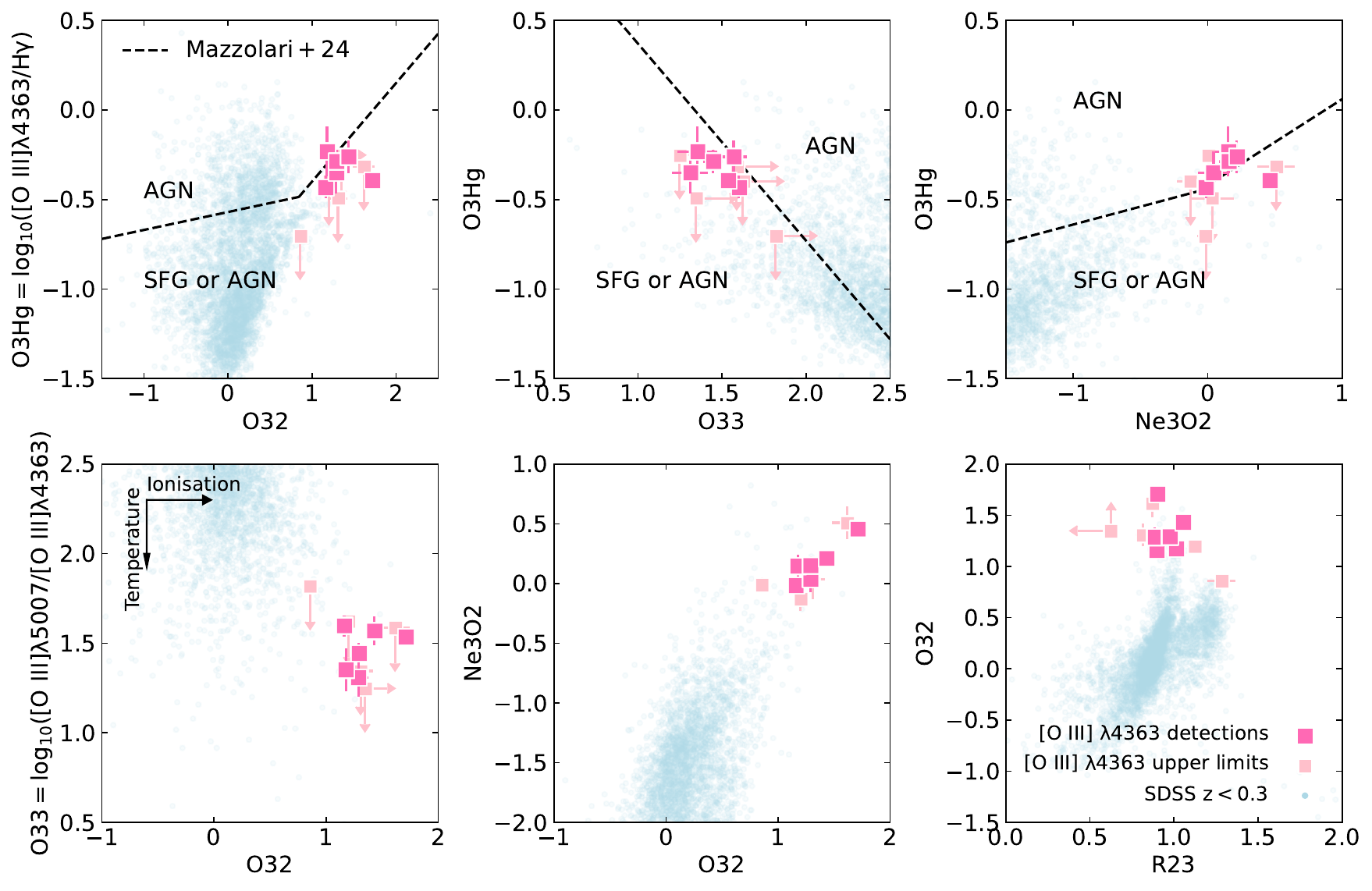}
    \caption{Line flux ratios of the $z=9$ sample, with [\oiii]$\lambda4363$ detections in hot pink, and $3\sigma$ upper limits in pale pink. The blue points show $z\approx0$ galaxies with [\oiii]$\lambda4363$ detections from from SDSS-DR15, for comparison. The dashed lines represent AGN diagnostics utilising the [\oiii] auroral line from \cite{Mazzolari24}, calibrated using local and high-z samples of SFGs and AGN.}
    \label{fig:linecomp}
\end{figure*}

Figure~\ref{fig:linecomp} shows the main line ratios of the strongest nebular lines, highlighting the typical ionisation fields, electron temperatures and diagnostics to separate active galactic nuclei (AGN) vs. star-forming galaxies. Here we also compare to the low-redshift ($z < 0.3$) galaxy sample from the Sloan Digital Sky Survey \citep[SDSS;][]{Aihara11}, for reference. The SDSS line fluxes were corrected for dust attenuation using the Balmer decrement, assuming $\rm H\alpha/H\beta = 2.86$ (Case B, $T_e=10,000$K), and a \cite{Calzetti2000} dust curve. 

Strong nebular line ratios have typically been used to distinguish the intrinsic emission from AGN or star-forming regions. The classic Baldwin-Phillips-Terlevich (BPT) diagram \citep{Baldwin81} separating the dominant emission components is based on the [\oiii]\,$\lambda 5007$/H$\beta$ vs. [\nii]\,$\lambda 6584$/H$\alpha$ line ratios. However, this is obfuscated at high redshifts due to the general increased intensity of the ISM ionisation fields \citep{Calabro24,RobertsBorsani24}, and we are unable to measure [\nii]\,$\lambda 6584$/H$\alpha$ with NIRSpec beyond $z\approx8.3$. New classification diagnostics have therefore been developed, focusing on particularly the O3Hg = $\rm log_{10}$([\oiii]\,$\lambda 4363$/H$\gamma$) line ratio \citep{Mazzolari24,Backhaus25}. In Panels (a)-(b) of Fig.~\ref{fig:linecomp}, we show the O3Hg line ratios compared to O32 = $\rm log_{10}$([\oiii]\,$\lambda 5007$/[\oii]\,$\lambda 3727$) and O33 = $\rm log_{10}$([\oiii]\,$\lambda 5007$/[\oiii]\,$\lambda 4363$), representing the ionisation field and $T_e$, respectively. We note that the O3Hg line ratios are only slightly higher than the low-redshift reference sample, likely due to a combination of higher $T_e$ but lower metallicities. While the majority of our sample are located in the star-forming locus for panels (a)-(b), they lie close to the boundary, and many are situated in the AGN region (albeit also along the boundary) for panel (c), where O3Hg is plotted with Ne3O2 = $\rm log_{10}$([\neiii]\,$\lambda 3869$/[\oii]\,$\lambda 3727$). Given the difficulty of distinguishing star-forming galaxies and AGN at high-z, and relying on faint emission lines for the diagnostics, we cannot rule out the presence of AGN in the sample. If present, the contribution could affect metallicity and stellar mass measurements. However, since the sample lies along the boundary, and there is no overwhelming evidence to suggest AGN are present, we assume these are star-forming sources in the following analysis. 

In Panel (d), we show O32 relative to the O33, the latter inversely correlating the electron temperature $T_e$ of the \hii\ region. We find no clear correlation between the lines ratios for the galaxies at $z=9-10$ but highlight the typical lower [\oiii]\,$\lambda 5007$/[\oiii]\,$\lambda 4363$ ratios at high-$z$, indicating higher $T_e$ in the ISM of these galaxies. In Panel (e), we show the Ne3O2 vs the O32 line ratios, which are both sensitive to the same ionisation field since [\neiii] and [\oiii] originate in the same high-ionisation zone of the \hii\ region (with ionisation potentials greater than $\sim$ 40 and 35 eV). We find that both these line ratios are substantially higher than for the typical low-redshift galaxy population, as also found previously for \jwst-observed galaxies up to $z\sim 8$ \citep{Schaerer22,Curti23a,RobertsBorsani24}. This indicates a higher ionisation parameter in these early systems, with median ${\rm log}U \approx -2$, quantified from diagnostics based on O32 and Ne3O2 \citep{Diaz2000, Witstok21}. 

\subsection{Direct $T_e$-based metallicities} \label{ssec:logoh}

We determine the metallicities using the direct `$T_e$-based' method \citep{Peimbert17, PerezMontero17, Kewley19, Osterbrock06} which relies on measuring the electron temperature $T_e$ to determine gas-phase metallicity 12+log(O/H). This assumes low electron densities ($n_e \lesssim 10^{5}$\,cm$^{-3}$), which appear common for galaxies at $z\sim 10$ \citep{Hsiao24a, Isobe23, Topping25}. %We use both PyNeb and the Izotov+06 iterations.

We estimate electron temperature $T_e$ with the [\oiii]$\lambda4363$/[\oiii]$\lambda5007$ ratio, using the {\tt getTemDen} routine in {\sc PyNeb} \citep{Luridiana15}. As the [\oii] doublet is unresolved we do not have a direct constrain on electron density. However, $n_e = 300\, \mathrm{cm^{-3}}$ is typical of star-forming galaxies at $z\sim2.3$ \citep{Sanders16}; although \jwst\ results have shown that there is likely to be a trend of increasing electron density with redshift \citep{Topping25,Abdurrouf24,Li25,Isobe23,Harikane25}. While we fiducially assume an electron density of $n_e = 300\, \mathrm{cm^{-3}}$, we find changing the density up to $n_e \sim 1\times10^4\, \mathrm{cm^{-3}}$ has at most $0.02\,\rm dex$ impact on the abundance measurements. \cite{Harikane25} have suggested a more accurate method would be combining \jwst\ and \alma\ observations and using a 2-zone ISM model to trace both high and low-density regions, as assuming a constant density and temperature throughout the ISM may result in underestimating the oxygen abundance by up to 0.8 dex. 

As determining singly-ionised oxygen abundance $\rm O^+$ requires a temperature that traces the low-ionisation zone of the gas, we assume a relation between temperatures of the high and low ionisation zones $T_e([\mathrm{\oii}]) = 0.7 \times T_e([\mathrm{\oiii}]) + 3000 \mathrm{K}$ from \citet{Garnett92}. \cite{Cataldi25} suggest a new $T_3-T_2$ relation for $z\sim2-3$, although using this calibration changes the oxygen abundance of the sample on average 0.01 dex, as the galaxies are highly ionised so the $\rm O^+$ abundance is generally low. 
The total oxygen abundance is then calculated by summing the contributions from different ionic abundances $\rm O/H^+ = O^+/H^+ + O^{++}/H^+$ obtained with {\tt getIonAbundance}. The contribution from $\rm O^{+++}$ is not included as it is likely to be negligible even under the most extreme ISM conditions \citep{Cullen25, Berg21}. The errors on all gas-phase properties are obtained using 1000 Monte Carlo simulations, where line fluxes are perturbed randomly according to their measured uncertainties. We employ the transition probabilities and collisional strengths included in {\sc PyNeb}; all atomic data from \cite{FroeseFischer04}, collisional data from \cite{Kisielius09} for $\rm O^+$, and \cite{Aggarwal99} for $\rm O^{++}$.

\begin{figure*}
    \centering
    \includegraphics[width=1.0\linewidth]{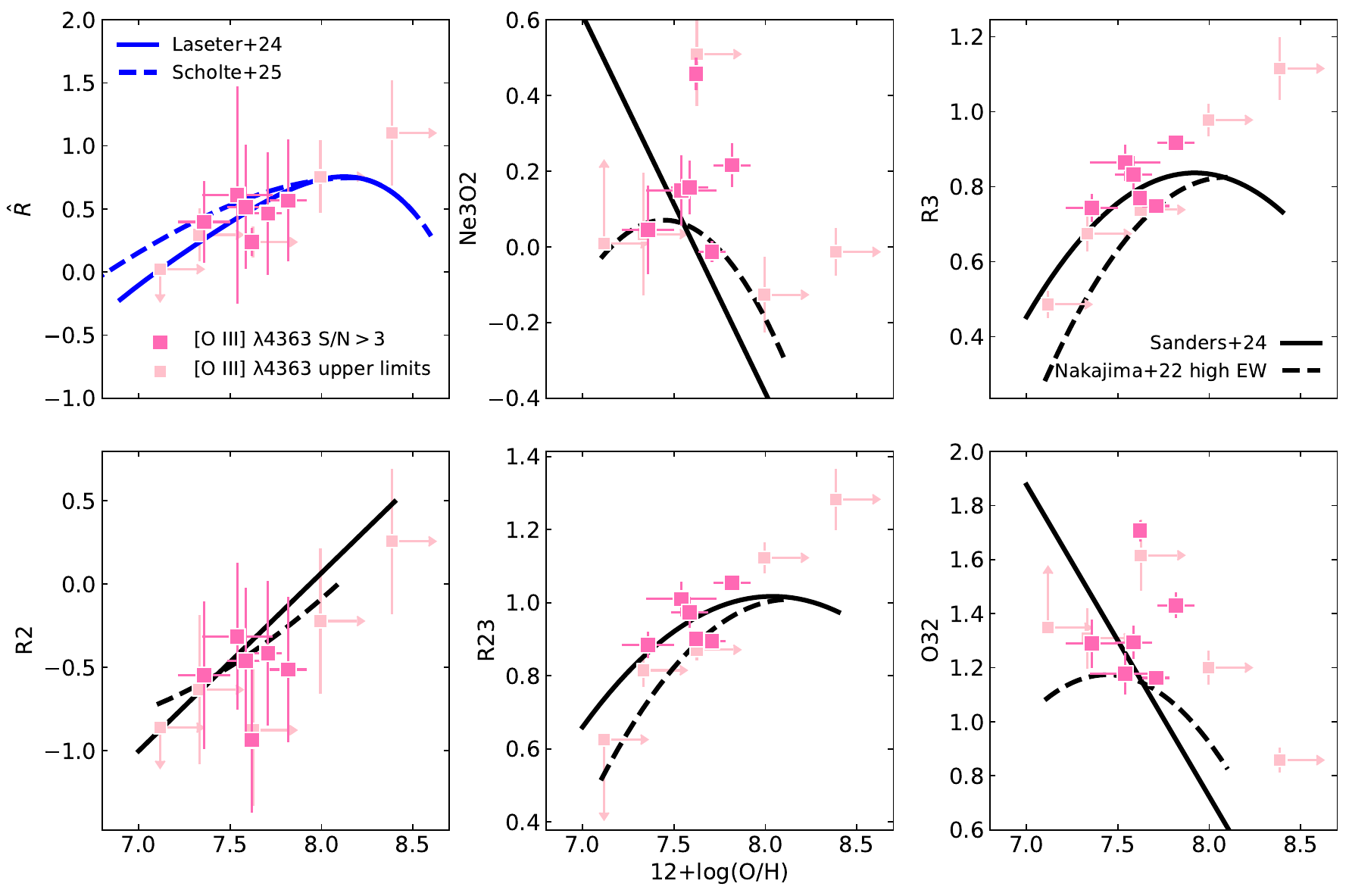}
    \caption{$T_e$-based metallicities and strong-line ratios for the $z>9$ sample. Direct oxygen abundances are represented by hot pink square markers, with $3\sigma$ upper limits in pale pink. We compare to various literature strong-line diagnostics: the solid and dashed blue relations shows \cite{Laseter2024} and \cite{Scholte25} $\hat{R}$ calibrations, solid and dashed black curves are $\rm Ne3O2, \ R3, \ R2, \ R23, \ O32$ diagnostics from \cite{Sanders24} and \cite{Nakajima22} respectively.}
    \label{fig:emlinediag}
\end{figure*}

The determine direct metallicities (and 3$\sigma$ upper limits) for the full sample are listed in Table~\ref{tab:props} and shown in Fig.~\ref{fig:emlinediag}, ranging from 12+log(O/H) $\approx 7.1 - 8.3$. We compare to high-$z$ literature diagnostics from \cite{Sanders24}, \cite{Laseter2024}, \cite{Scholte25}, and high equivalent width (EW) low-z diagnostics from \cite{Nakajima22}. Our sample are generally in good agreement with these calibrations, (with considerable scatter, as was also seen in previous works) though we do not have a large enough statistical sample to distinguish between those that diverge. Our derived metallicities for RXJ2129-11767-11027 and JADES-GS-265801 are consistent within $\sim0.1$dex to the values reported in previous literature; respectively 12+log(O/H) = $7.48 \pm 0.08$ \citep[][from strong-line ratios]{Williams23} and $7.49 \pm 0.11$ \citep[][using the direct method, with ionisation correction factors]{Curti25}. 

% \subsection{Star-formation rates}

\subsection{Spectro-photometric SED modelling} \label{sec:bagpipes}

%{\bf Rashmi check and expand/update}
To derive the stellar continuum properties of our sample, in particular the stellar masses, we perform spectro-photometric modelling of the spectral energy distribution (SED) of each source based on the \jwst/NIRSpec PRISM spectroscopy and available broad-band NIRCam photometry. We summarise our modelling methods below.%; a more detailed description will be presented in Gottumukkala et al. (in prep).

In order to estimate the SED model posteriors, we use the code Bayesian Analysis of Galaxies for Physical Inference and Parameter EStimation \citep[{\sc Bagpipes},][]{Carnall2018,Carnall19} which performs nested-sampling using the {\sc Nautilus} \citep[][]{Lange2023} algorithm. Within {\sc Bagpipes}, we use the Binary Population and Spectral Synthesis code (BPASS) v2.2.1\footnote{\protect\url{https://bpass.auckland.ac.nz/9.html}} stellar population synthesis (SPS) models \citep{Stanway2018} to build galaxy SEDs based on a broken power-law initial mass function (IMF), where the slope of the power law is $\alpha_1=-1.30$ for ${M_\star \in (0.1,0.5)\ M_\odot}$ and $\alpha_2=-2.35$ for ${M_\star \in (0.5,300)\ M_\odot}$. We use nebular line and continuum emission models from the {\sc Cloudy} photoionisation code \citep[][]{Ferland2017, Byler2017}, where the {\sc Cloudy} grids are calculated using the BPASS SPS models as the input ionising source.
%, an IGM attenuation model from \cite{Inoue2014} 
%and user-specified star-formation histories (SFHs) and dust laws. The software performs nested-sampling using the {\sc Nautilus} \citep[][]{Lange2023} algorithm. In this work, we use the Binary Population and Spectral Synthesis code (BPASS) version 2.2\footnote{\protect\url{https://bpass.auckland.ac.nz/9.html}} SPS models, and accordingly compute {\sc Cloudy} nebular line and continuum emission models.

To account for slit losses, the \jwst/NIRSpec PRISM spectra are flux-calibrated to the NIRCam broadband photometry self-consistently within the {\sc Bagpipes} framework with a second-order Chebyshev polynomial. Gaussian priors are set on the polynomial coefficients by performing an initial least-squares fit to the ratio between the observed photometry to the synthetic photometry (calculated by integrating the NIRSpec spectrum through the NIRcam filter profiles\footnote{\protect\url{https://svo2.cab.inta-csic.es/theory/fps/}}). From an initial SED modelling run, we find that only two sources in our sample require photometric scaling, JADES-GN-3990 and UNCOVER-2561-3686; for the rest, we perform SED modelling without the calibration described above.

We model the SEDs with a non-parametric continuity star-formation history \citep[SFH;][]{Leja19}, which allows for flexible star-formation rates (SFRs) in a sequence of user-defined time steps. The initial few time step edges are fixed in lookback time at [0, 10, 50, 100] Myr, after which the timesteps are equally space in logarithmic lookback time between 100 Myr and the age on the universe at $z=30$. The number of time steps therefore is redshift-dependent. We impose a Student's {\sl t} distribution on the change in the SFR between adjacent time bins, with the distribution being centred at 0 with a scale factor $\sigma = 0.3$ and $\nu = 2$ \citep{Leja19}. We set a logarithmic prior on the formed stellar mass, ${M_\text{formed} / M_\odot \in (10^{6}, 10^{14})}$. We set a Gaussian prior on the metallicity, centred on the direct $T_e$-based measurements derived in Sect.~\ref{ssec:logoh} with a scatter of 10\% ${Z_\odot}$. For galaxies with only a lower limit on $T_e$-based metallicities, strong-line diagnostics were instead used to quantify the oxygen abundance (see Eq.~\ref{eq:chi2}; Sect.~\ref{ssec:MZR}). %This mainly drives the stellar mass estimates, which are, however, mostly consistent when leaving the metallicity as a free parameter (posteriors centre around $\approx 10\%$ ${\rm Z_\odot}$).

The dust is modelled with a \cite{Salim18} dust curve, where we fix the slope of the attenuation curve to $\delta = -0.3$ and the strength of the 2175\AA\ UV bump is fit in the range $B\in(0,3)$ with a uniform prior. An extra factor of attenuation is applied to starlight within stellar birth clouds, where $\eta=2$. We adopt a uniform prior on the $V$-band attenuation, where ${A_V \in (0, 1)\ {\rm mag}}$.

The nebular ionisation parameter is varied in the range ${\log U \in (-4, -1)}$ with a uniform prior. We set a tight Gaussian prior on the redshift centered at the spectroscopic redshift derived from {\sc msaexp} on the DJA, with $\sigma_z = {0.001\times(1+z_{\rm spec})}$. The velocity dispersion is fixed to 100 km/s. A logarithmic prior is set on the white noise scaling in the range (1.,10.).

% {\bf Rashmi update with your awesome stuff here:} To improve the emission-line modelling in {\sc Bagpipes}, we do X and Y. From some initial modelling, we find that {\sc Bagpipes} is often unable to reproduce the observed strengths and line ratios of certain emission lines, which leads to poor fits of the lines themselves and the continuum regions around the lines. To mitigate this effect, we model the residual between the data and the {\sc Bagpipes} SED with line components centered at the observed wavelengths of the [\oiii]\,$\lambda\lambda 4959,5007$ doublet; this model is then added to the modelled SED at the stage of the log-likelihood calculation. In this way, we ensure that that continuum drives the fit, and not the emission lines, thereby giving more reliable continuum properties such as stellar masses.

%An example of the best-fit SED model from {\sc Bagpipes} is shown in Fig.~\ref{fig:linefitex}. \textcolor{red}{CP to update Fig 1 with final bagpipes fit.}
We caution that the posterior parameter estimates from the SED modelling are heavily model-dependent, where in particular the stellar mass may be subject to `outshining effects' \citep[e.g.,][]{GimenezArteaga23,Narayanan24}, and the assumed SFH and IMF \citep[e.g.,][]{Carnall2018,Leja19,Steinhardt23,Strait23}.

\section{Galaxy properties at $z\approx 10$} \label{sec:gal}

\subsection{Rest-frame UV properties}

The derived galaxy properties are summarized in Table~\ref{tab:props}. We first consider the direct, measured properties of the targeted galaxies at $z=9-10$.
%In Figure~\ref{fig:muvbeta}, we show the absolute rest-frame UV magnitude ($M_{\rm UV}$) as a function of the UV spectral slope, $\beta_{\rm UV}$, measured directly from the spectrum. These properties generally represent the contribution of massive stars to the observed stellar light and a combination of the average age of the stellar populations and the overall dust reddening.
For the absolute rest-frame UV magnitude, we integrate the flux density of the spectrum with a $100\AA$ top-hat filter centred at rest-frame $1500\,\AA$. We fit a standard power-law slope, assuming $F_{\rm \lambda, obs} \propto \lambda^{-\beta_{\rm UV}}$ to determine $\beta_{\rm UV}$, carefully masking the most prominent nebular emission lines in the modelling. The large scatter, from $\beta_{\rm UV} = -1.5$ to $-2.8$ and $M_{\rm UV} = -16$ to $-21$\,mag are consistent with larger samples using photometric \citep{Topping24, Cullen24, Austin24} and spectroscopic \citep{Saxena24, Dottorini24} measurements. We do not observe any strong correlations between $M_{\rm UV}$ and $\beta_{\rm UV}$ in our spectroscopic sample at $z=9-10$, where typically brighter, more massive galaxies would be expected to show redder rest-frame UV slopes.

% Write your own, nice-looking table
%\begingroup % Optional: affects only this particular table
%    \setlength{\tabcolsep}{10pt} % Default value: 6pt
    \begin{table*}
     \renewcommand{\arraystretch}{1.0} % Default value: 1
    %    \centering
                \caption{Summary of the main galaxy properties for our $z>9$ sample, including UV magnitude $M_{UV}$, UV continuum slope $\beta_{UV}$, stellar mass, star-formation rates derived from H$\beta$, electron temperature $T_e$, and direct $T_e$ based metallicities derived using {\sc PyNeb}. }
        \begin{tabular}{ l c c c c c c c c c c }  
                    \hline \hline    
            Source & $M_{\rm UV} / {\rm mag}$ & $\beta_{\rm UV}$  & $\log_{10} (M_\star / M_\odot) $ & $\log_{10} ({\rm SFR}_{\rm H\beta} / M_\odot\,{\rm yr}^{-1})$ & 12+log(O/H) & $T_e / {\rm 10^4 K}$  \\
            \hline    
            CAPERS-EGS-25297 & $-19.47 \pm 0.26$ & $-1.48 \pm 0.05$ & $8.39^{+0.04}_{-0.02}$ & $1.32\pm 0.02$ & $7.69 \pm 0.08$ & $1.7 + 0.12 $ \\
            UNCOVER-2561-13151 $^{(*)}$ & $-16.29 \pm 0.27$ & $-2.45 \pm 0.06$ & $6.97^{+0.10}_{-0.10}$ & $-0.48\pm0.03$ & $>7.65$ & $<1.71 $ \\
            JADES-GN-55757 &$-19.19 \pm 0.51$ &  $-2.34 \pm 0.09$ & $8.05^{+0.16}_{-0.17}$ & $0.74\pm0.04$ & $>7.35$ & $<2.41$ \\
            UNCOVER-2561-22223 $^{(*)}$ & $-17.15 \pm 0.55$ &$-2.60 \pm 0.19$ &  $6.99^{+0.15}_{-0.05}$ & $0.10\pm0.03$ & $>7.11$ & $< 2.91  $     \\   
            RXJ2129-2767-11027 $^{(*)}$ & $-15.86 \pm 0.75$ & $-1.43 \pm 0.19$ & $6.87^{+0.04}_{-0.03}$ & $-0.19\pm0.04$ & $7.36\pm 0.14$ & $2.58 \pm  0.52$ \\
            JADES-GS-265801 & $-19.68 \pm 0.07$ & $-2.50 \pm 0.01$ & $8.30^{+0.03}_{-0.03}$ & $0.75\pm0.01$ & $7.61\pm 0.04$ & $1.83 \pm  0.07$\\  
            CAPERS-EGS-87132 & $-18.45 \pm 0.63$ & $-1.83 \pm 0.13$ & $8.19^{+0.13}_{-0.15}$ & $0.48\pm0.05$ & $7.53\pm 0.17$ & $2.39 \pm  0.49$ \\
            JADES-GN-3990 & $-18.65 \pm 0.39$ & $-1.30 \pm 0.04$ & $8.59^{+0.18}_{-0.12}$ & $1.23\pm0.04$ & $7.57\pm 0.11$ & $2.09 \pm 0.18$\\   
            UNCOVER-2561-3686 $^{(*)}$ & $-20.56 \pm 0.18$ & $-2.08 \pm 0.03$ & $9.41^{+0.01}_{-0.02}$ & $0.50\pm0.08$ & $>8.34$ & $<1.35$ \\
            RUBIES-UDS-833482 & $-19.69 \pm 0.84$ & $-2.24 \pm 0.14$ & $8.39^{+0.03}_{-0.02}$ & $1.14\pm0.04$ & $>7.94$ & $<1.66$\\   
            CAPERS-UDS-22431 & $-19.76 \pm 0.22$ & $-2.58 \pm 0.04$ & $8.30^{+0.04}_{-0.04}$ & $0.77\pm0.02$ & $7.81\pm 0.11$ & $1.76 \pm  0.18$\\
            \hline \hline  
        \end{tabular}  \\
        {\bf Notes.} $^{(*)}$ Lensed galaxies, with reported properties corrected for magnification with magnification factors for RXJ2129-2767-11027 obtained from \cite{Williams23}, and for UNCOVER galaxies from the lensing magnification online catalogue \footnote{\url{https://jwst-uncover.github.io}} \citep{Price25, Bezanson24, Furtak23}.
        \label{tab:props}
    \end{table*}
%\endgroup
\footnotetext{\url{https://jwst-uncover.github.io}}

%\begin{figure}
%    \centering
%    \includegraphics[width=8.8cm]{Figs/MUV_betaUV_litcomp.png}
%    \caption{UV continuum slope $\beta_{UV}$ versus UV magnitude $M_{UV}$ for the galaxies in our $z=9$ sample. We compare to literature relations from photometric catalogues \citep{Topping24, Cullen24}, showing brighter galaxies have redder slopes. We observe similar high scatter to the photometrically derived calibrations.}
%    \label{fig:muvbeta}
%\end{figure}

\begin{figure}\label{fig:ohmuv}
    \centering
    \includegraphics[width=8.8cm]{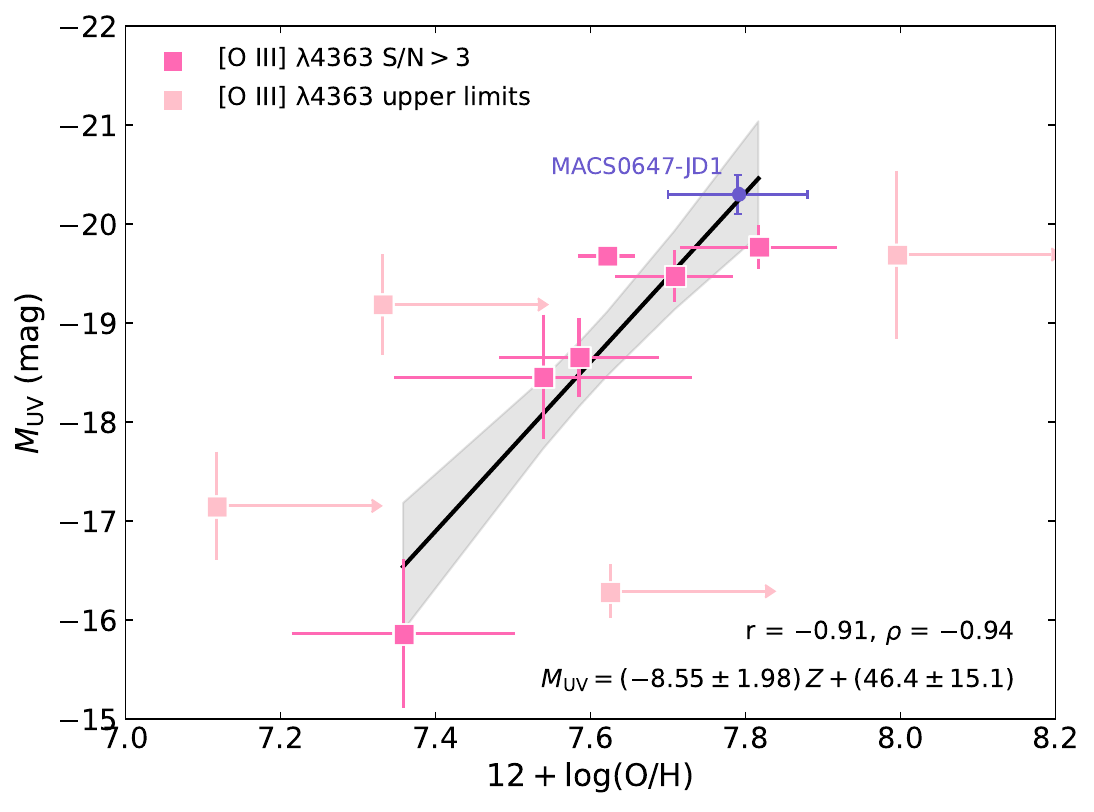}
    \caption{UV magnitude $M_{UV}$ versus direct $T_e$-based metallicity 12+log(O/H). Also shown is the direct metallicity measurement of MACS0647-JD1 at $z=10.17$ from \cite{Hsiao23}. We fit an empirical relation to the measurements (including MACS0647-JD1, but not the 3$\sigma$ upper limits), shown in solid black with 1$\sigma$ errors represented by the grey shaded region. The best-fit relation is reported in the bottom-right with $Z={\rm 12+log(O/H)}$, with the associated Spearman and Pearson correlation coefficients, $\rho$ and $r$, indicating a  strong correlation; the most massive, UV-bright galaxies are more metal enriched.}
    %The best-fit slope ${\rm d}M_{\rm UV}/{\rm d}Z = -8.29 \pm 2.17$, with a Spearman correlation coefficient of $\rho = -0.83$ and a Pearson correlation coefficient of $r=-0.89$, shows a strong correlation; the most massive, UV-bright galaxies are more metal enriched.}
    \label{fig:ohmuvbeta}
\end{figure}

In Figure~\ref{fig:ohmuv}, we compare $M_{\rm UV}$ to the oxygen abundance $12+\log$(O/H). For comparison to our sample we plot the lensed galaxy MACS0647-JD1 at $z=10.17$ \citep{Hsiao23}, as the only other galaxy with a direct metallicity determination at $z\sim10$. We find a clear correlation between $M_{\rm UV}$ and direct $12+\log$(O/H) (including MACS0647-JD1, but not lower limits), with a best-fit relation of $M_{\rm UV} = (-8.55 \pm 1.98)\,Z + (46.4 \pm 15.1)$, with $Z = {\rm 12+log(O/H)}$. This is expected given that the most massive, star-forming galaxies are likely also to show higher metal enrichment at a given SFR. This strong correlation (with a Spearman correlation coefficient $\rho = -0.94$, Pearson coefficient $r=-0.91$) further provides a viable relation to infer the gas-phase metallicities for galaxies at $z\gtrsim 10$ with only measures of $M_{\rm UV}$. We do not recover any apparent trend relating $\beta_{\rm UV}$ to $12+\log$(O/H), where increased dust reddening would be expected to scale with the metallicity. This indicates that the rest-frame UV slope is likely not representing the overall dust content of high-redshift galaxies, potentially due to more dominating nebular continua at low metallicities \citep{Cameron24,Katz25}. % Too strong? 

\subsection{Star-formation rates and timescales}

The relationship between the star-formation rate (SFR) and stellar mass, $M_\star$, also known as the star-forming galaxy main sequence (SFMS), is tightly connected for star-forming galaxies, though varying in normalization across redshifts \citep[e.g.,][]{Brinchmann04,Daddi07,Whitaker12,Speagle14,Lee15,Thorne21}. This indicates an evolving specific SFR (sSFR = SFR/$M_\star$), with galaxies at higher redshifts being more actively forming stars at a given stellar mass \citep[e.g.,][]{Topping22}. To investigate this relation at $z\approx 10$, we here adopt $M_\star$ inferred from the non-parametric SFH model outlined in Sect.~\ref{sec:bagpipes} and derive the SFR based on the H$\beta$ Balmer line luminosity. This uniquely traces the on-going star formation on relatively short, $\sim 10$\,Myr timescales, compared to standard UV- or SED-based estimates which trace SFRs over $\sim 100$\,Myr. We use the prescription from \citet{Shapley23} to determine ${\rm SFR_{H\beta}}$ based on H$\alpha$, assuming a constant ratio of ${\rm H\alpha / H\beta} = 2.76$ dictated from the Case B recombination scenario at $T_e = 2\times10^{4}$\,K \citep{Osterbrock06} and the average metallicity for our sample of $\approx 10\%$ Solar. 

\begin{figure}
    \centering
    \includegraphics[width=9cm]{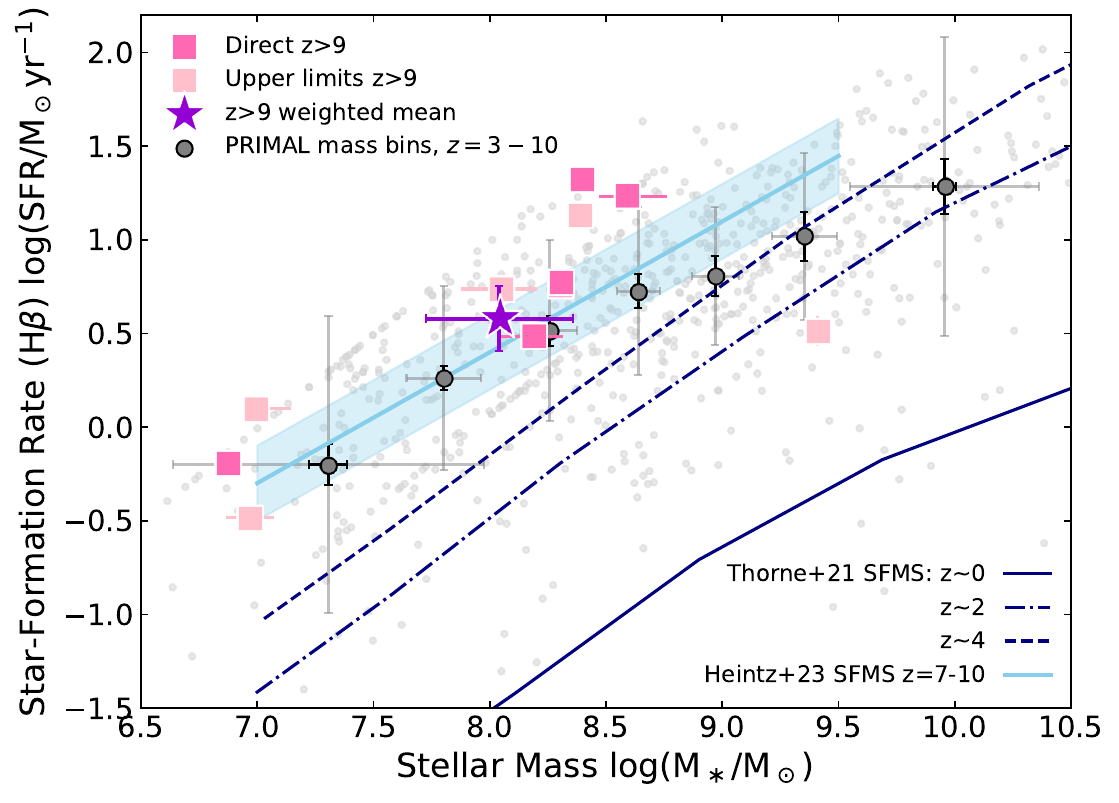}
    \caption{The star-forming galaxy ${\rm SFR}-M_\star$ main-sequence, the $z>9$ galaxies are plotted in pink, with the weighted mean of the sample represented by the purple star. To compare, we plot the PRIMAL sample at $z=3-10$ as grey points, with the median of stellar mass bins indicated by grey markers. The solid black error bars on these markers represent the error on the median, with grey error bars showing the standard deviation of the bin. We also plot the linear relation (and empirical scatter) of the best-fit SFMS for $z=7-10$ from \cite{Heintz23_FMR} in light blue, with dark blue curves showing the evolution of the best-fit main sequence for $z\sim 0, 2, 4$ from \cite{Thorne21}.}
    \label{fig:sfms}
\end{figure}

In Figure~\ref{fig:sfms}, we show the ${\rm SFR_{H\beta}}$-$M_\star$ relation for the galaxies at $z=9-10$, compared to the full \jwst-PRIMAL sample at $z=3-9$ \citep{Heintz25} and other recent literature measurements at $z\gtrsim 6$ \citep{Heintz23_FMR,Shapley23}. While the sample clearly show elevated SFRs at a given mass compared to more local estimates for galaxies at $z=0-4$ \citep[e.g.][]{Thorne21}, the SFMS is overall consistent in normalization and slope as the other literature measurements at $z\gtrsim 6$, potentially with a mild increase in SFR of 0.15\,dex on average. 
This indicates only a marginal evolution in the SFMS from $\approx 1$\,Gyr to 500\,Myr, consistent with the overall observed trend with redshift. The median sSFR is $58$\,Gyr$^{-1}$, $3-8\times$ the average value found for massive, UV bright galaxies at $z=6-8$ \citep{Topping22}. These results suggest that SFMS is already in place at $z\approx 10$, just 470\,Myr after the Big Bang, and that our sample is generally comprised of `typical', but very active star-forming galaxies during this epoch.

We now investigate the SFR timescales for the sample galaxies at $z\approx 10$, searching for potential signs of short-lived bursts of star formation, which might be related to the observed overabundance of UV bright galaxies at $z\gtrsim 10$ \citep[e.g.,][]{Sun23,Gelli24}. We compare the SFRs derived from H$\beta$ and the best-fit SED for the full galaxy sample at $z=9-10$. We find that, on average, the SFR on 10\,Myr to 100\,Myr timescale is $\log_{10}({\rm SFR_{10}/SFR_{100}}) = 0.4$, which is $\gtrsim 2\times$ higher than the average value found at $z\approx 6$ \citep{Endsley24,Cole25}, but consistent with individual galaxy studies at $z\gtrsim 10$ \citep{Kokorev25}. These measurements thus provide strong evidence for highly stochastic or more bursty star formation in the general population of star-forming galaxies at $z\gtrsim 10$. 
%SFR on the 10\,Myr timescale is, on average, $3\times$ higher than the SFR derived over the past 100\,Myr. This is 
% SFRline / SFRsed vs stellar mass. 

%\begin{figure}
%    \centering
%    \includegraphics[width=9cm]{Figs/SFR_SED_Hb.png}
%    \caption{Add plot with SFRhbeta vs SFRsed vs. redshift (!) or mass?.}
%    \label{fig:sfrlinesed}
%\end{figure}

\subsection{Mass-metallicity relation at $z\approx 10$}\label{ssec:MZR}

\begin{figure}
    \centering
    \includegraphics[width=8.8cm]{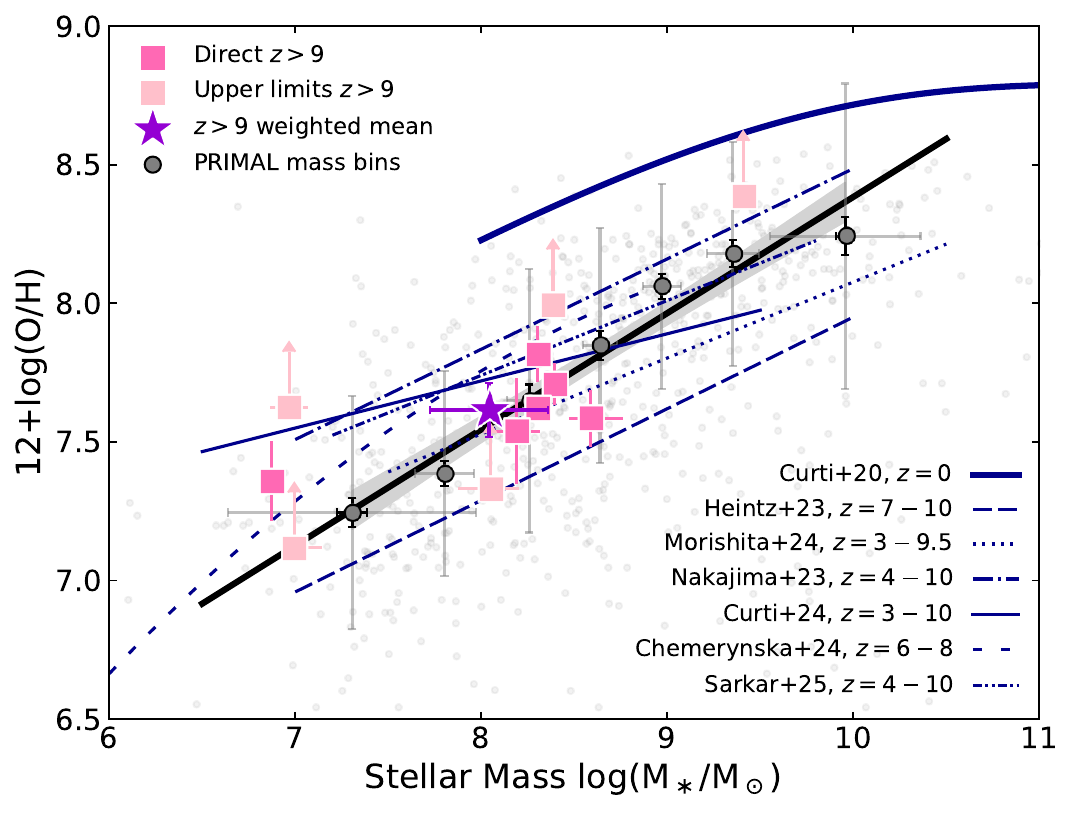}
    \caption{Mass-metallicity relation at $z>9$. Direct $T_e$ metallicities and lower limits for the $z>9$ sample are shown in hot pink and light pink respectively, with the weighted mean and 1$\sigma$ error of the sample indicated by the purple star. Various mass-metallicity relations from literature are plotted \citep{Heintz23_FMR, Curti20, Morishita24, Nakajima23, Curti24, Chemerynska24b}. Light grey scatter points represent the PRIMAL $z=3-10$ sample, with grey circular markers and their associated solid black error bars being the median and 1$\sigma$ error of stellar mass bins. The grey error bars represent the standard deviation of mass bins. The bold solid black line shows the best fit to the PRIMAL stellar-mass bins.}
    %, with a best-fit relation ${\rm 12 + log(O/H)} = (0.42 \pm 0.04)\, \log(M_\star) + (4.19 \pm 0.34)$.}
    \label{fig:massmet}
\end{figure}

We then consider the mass-metallicity relation (MZR) for our sample in Fig.~\ref{fig:massmet}, extending previous analyses with \jwst\ out to $z>9$.  
The scaling relation between stellar mass and gas-phase metallicity is among the most studied galaxy relationships, with a relatively small intrinsic scatter ($\sim 0.1$ dex) and strong correlation, showing more massive systems become more metal-enriched. The shape of the local MZR shows a steeper slope at low masses, which flattens at higher masses, with a turnover $\rm \sim 10^{10}M_\odot$ \citep[see e.g.][for a review]{Maiolino19}. %Physical processes behind shape?
Previously, detections of [\oiii]\,$\lambda 4363$ at $z>3$ were mostly inaccessible, but thanks to the sensitivity and wavelength coverage of NIRSpec, and the increased sample size of direct metallicity measurements, early \jwst\ studies of the MZR suggest the relationship is already in place at $z=6-8$ \citep[e.g.,][]{Curti24, Nakajima23}.
We compare our direct $T_e$ measurements and lower limits to various high-$z$ MZR relations from literature, which use both the direct $T_e$ method \citep{Curti24, Nakajima23, Morishita24} and strong-line calibrations \citep{Heintz23_FMR}. As expected, our sample galaxies are similarly less enriched for a given stellar mass than local galaxies \citep[e.g.,][]{Curti20}; between $\sim 0.5 - 1.0$ dex. 

As a benchmark, we also include additional high-$z$ galaxies from the \jwst-PRIMAL archival study \citep[][grey markers and scatter points]{Heintz25}, selected at $z>3$ and with H$\beta$ detections with S/N$>3$. The metallicities of these galaxies were obtained using strong-line calibrations from \cite{Sanders24} and \cite{Laseter2024}, using a joint $\chi^2$ approach taking into account each strong-line ratio available ($\hat{R}$, $\rm Ne3O2$, $\rm R3$, $\rm R2$, $\rm R23$, $\rm O32$), inversely weighted by its intrinsic scatter. 
\begin{equation}\label{eq:chi2}
    \chi^2 (x) = \sum_n \frac{\Bigl(R_{obs, n}-R_{cal, n}(x)\Bigl)^2}{\sigma^2_{obs, n}+\sigma^2_{cal, n}}
\end{equation}
Where $x = \mathrm{12+log(O/H)}$, $R_{obs}$ is the observed line ratio with error $\sigma_{obs}$, $R_{cal}$ is the modelled line ratio, and $\sigma_{cal}$ is the reported scatter of the diagnostic \citep{Curti20}.

We divide 735 galaxies into 7 stellar-mass bins of 105 galaxies each, and fit a simple log-linear relation to the bins (solid black line), with a best-fit relation of ${\rm 12 + log(O/H)} = (0.42 \pm 0.04)\, \log_{10}(M_\star) + (4.19 \pm 0.34)$. The weighted mean of our $T_e$-based sample (depicted by a purple star) lies in line with the derived relation for the PRIMAL galaxies. 

\begin{figure}
    \centering
    \includegraphics[width=8.8cm]{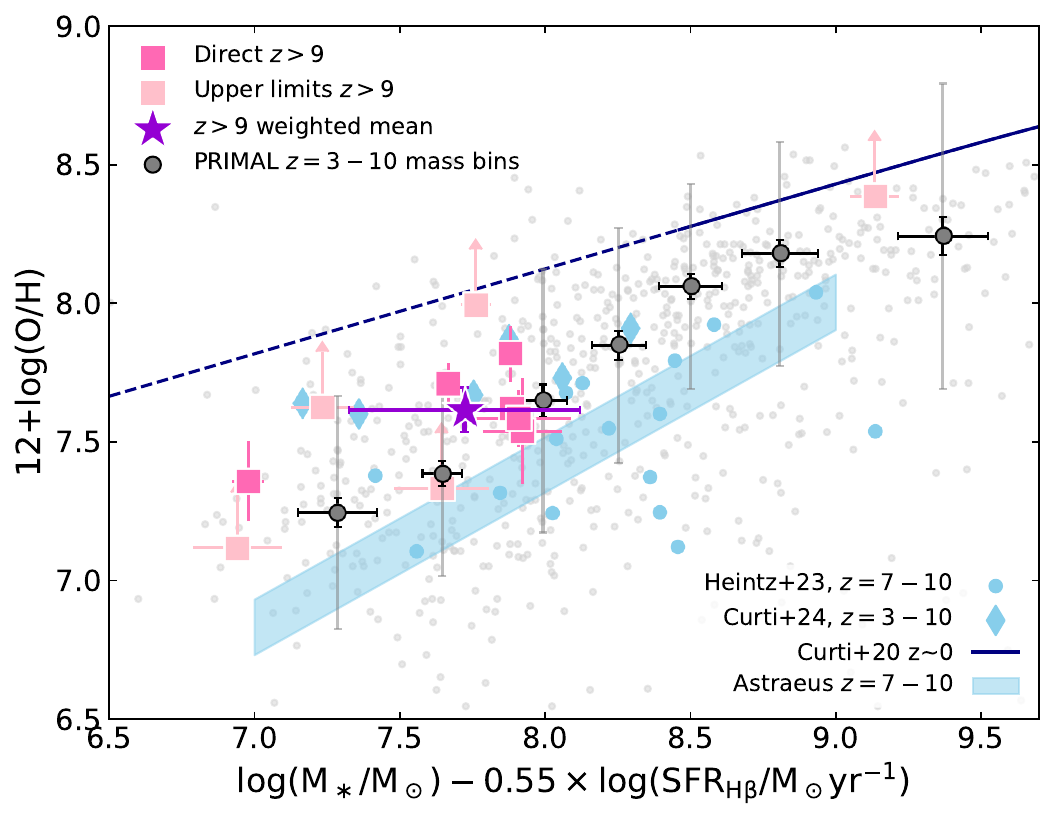}
    \caption{The Fundamental Metallicity Relation at $z>9$. Direct $T_e$ metallicities and upper limits for the $z>9$ sample are shown in hot pink and light pink respectively, with the weighted mean of the sample indicated by the purple star. The local FMR \citep{Curti20} is plotted in dark blue, and extrapolated to lower $\mu_\alpha$. Grey scatter points represent the PRIMAL $z=3-10$ sample, with grey markers being the median of stellar mass bins. The solid black and grey error bars on these points represent the error on the median, and standard deviation of each bin respectively. We compare to previous high-z \jwst\ results with diamonds from \cite{Curti24}, circles from \cite{Heintz23_FMR}. The blue shaded region shows predictions from the Astraeus simulations for $z=7-10$. }
    \label{fig:fundmet}
\end{figure}

%List - redshift, mass ranges, metallicity method for each literature. \cite{Curti20} z=0 with $T_e$ metallicities, logM = 8 - 11.5. \cite{Heintz23_FMR}, z=7-10, logM=7-10, strong-line metallicities. \cite{Nakajima23}, z=4-10, logM = 7-10, direct $T_e$ method. \cite{Morishita24}, z=3-9, logM = 7.5-10.5, direct $T_e$ method. 

\begin{figure*}
    \centering
    \includegraphics[width=15cm]{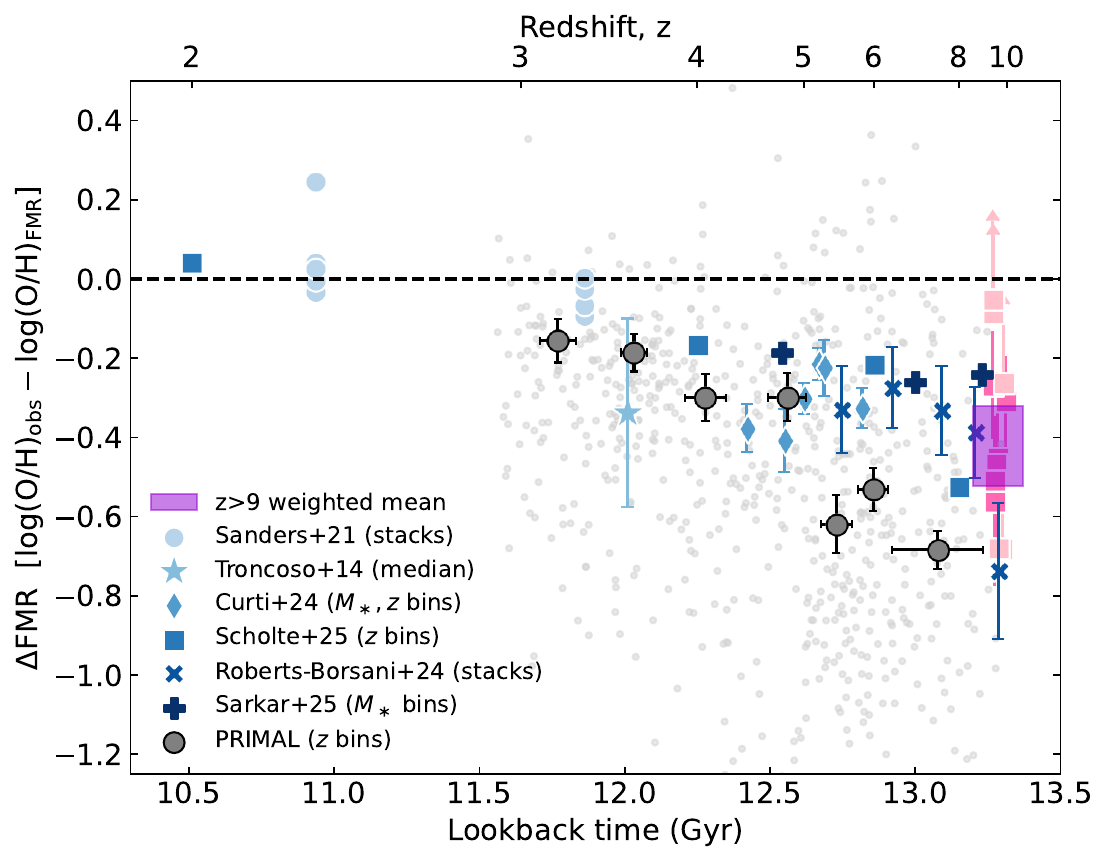}
    \caption{Offset from the FMR, plotted against lookback time and redshift, down to $z\sim2$. The weighted mean of $z>9$ direct-metallicity sample is shown by the purple box, with height and width spanning 1$\sigma$ error, and individual galaxies are plotted in pink. We compare to redshift bins of the PRIMAL $z=3-10$ sample, shown as grey circular markers, with errorbars representing the error on the bin medians. Other binned or stacked high-z ($z>2$) literature data are shown in different shades of blue, from \cite{Sanders21, Troncoso14, Curti24, Scholte25, RobertsBorsani24, Sarkar25}. There is a clear trend indicating that the offset from the FMR increases with redshift, with a systematic break somewhere between $2.5 < z < 3.0$. }
    \label{fig:fundmetz}
\end{figure*}

\subsection{Deviation from the fundamental-metallicity relation}
The MZR has been observed to have an additional dependence on the SFR, with highly star-forming galaxies appearing to be more metal-poor at a given stellar mass. To reduce the scatter on the MZR, \cite{Mannucci10} introduced the fundamental metallicity relation (FMR) with an additional parametrisation for SFR; 
\begin{equation}
    \mu_\alpha = \log_{10} M_\star - \alpha\times  \log_{10}({\rm SFR})
\end{equation}
Where $\alpha$ is a constant between 0 and 1, chosen to minimise the scatter in $\mu_\alpha -Z$ space \citep{Mannucci10,Andrews13}. The FMR appears universally constant out to $z\approx 3$ \citep{Sanders21}, though recent \jwst\ measurements report a break from this relation for the highest redshift galaxies \citep{Heintz23_FMR,Nakajima23,Curti24} with the exact transition period still debated (likely between $z\approx 4-8$). In Figure~\ref{fig:fundmet}, we compare our sample with early \jwst\ results from \cite{Heintz23_FMR}, and mass-redshift bins from \citep{Curti24}, as well as the predicted FMR from the {\tt Astraeus} simulation at $z=7-10$ \citep{Ucci23}. As background, we show the full PRIMAL sample at $z=3-10$ with metallicities based on the strong-line diagnostics. The weighted mean of our sample is 0.4\,dex below the extrapolated FMR for galaxies in the local Universe. It is generally consistent with the underlying PRIMAL high-redshift \jwst\ sample, that overall indicate a slightly steeper slope of $\alpha = 0.78$ than determined locally \citep[$\alpha = 0.32-0.65$;][]{Mannucci10,Andrews13,Curti20}.

%\cite{Mannucci10} reported that the FMR begins to deviate at $z\gtrsim 2.5$, which was later supported by observations \citep{Sanders20, Cresci19}, more?
%We compare our sample with early JWST results from \cite{Heintz23_FMR}, and mass-redshift bins from \citep{Curti24}, as well as the predicted FMR from Astraeus simulations. Our sample's weighted mean falls in line with the PRIMAL $z=3-10$ stellar mass bins (grey markers), $\sim 0.5$ dex offset. 

% KEH Moving this down to discussion
%The observed systematic offset at high-redshift could be explained by accretion of pristine neutral gas, although large scatter in the offsets may imply we are observing an era of galaxy formation where an equilibrium has not yet been established between star-formation, chemical enrichment, feedback, and inflows. 

In Figure~\ref{fig:fundmetz}, we compile the high-redshift observations and show the overall evolution of the FMR with redshift. We calculate the offset from the local FMR, using the parametrisation outlined in \cite{Curti20} (Equation 5; $\alpha=0.56$). As seen in Fig.~\ref{fig:fundmet}, the majority of the $z=9$ sample lie in a $M_\star$-SFR regime which the low-z galaxies do not cover, where the FMR has been extrapolated. However, local high-z analogue samples that probe lower $\mu_\alpha$ do appear to follow the expected FMR \citep[e.g.][follows \cite{Andrews13,Curti20}]{ArellanoCordova24}. We observe a systematic break from the local FMR for galaxies at $z\gtrsim 3$, potentially evolving from -0.25\,dex at $z=3$ down to -0.7\,dex at $z>6$. The weighted mean of the $z=9-10$ sample is offset by 0.4\,dex from the local relation, consistent with other \jwst\ studies \citep{Heintz23_FMR,Curti24, RobertsBorsani24, Scholte25}. These observations indicate substantially different physical properties driving galaxy growth within the first Gyr of cosmic time. We will interpret and discuss potential scenarios to explain the observed deviations from the FMR in more detail in Sect.~\ref{sec:disc} below.  

%\textit{Note there are different reported values of $\alpha$ depending on the sample, even at z=0 - Curti+20 0.55, Andrews+Martini13 0.65, Mannucci+10 0.32. Garcia+24 discusses the possibility of a redshift-evolving FMR, with simulated data showing evidence for an evolving relation with changing $\alpha_{min}$.
%A free fit to the stellar-mass bins for z=9 sample gives $\alpha = 0.78$}

% Plot of offset with redshift - PRIMAL z=3-10, Curti+24 mass bins, Scholte+25, Sanders+21, Troncoso+14, weighted mean. 

% Test this / colour-code against SFR density. 
% Derive SFR From H-beta and UV.
% First look into star-forming SFR vs Mstar main-sequence and find X. Note sure we should show a plot, but definitely interesting that the SFRs are higher by approx. 0.5 dex compared to the z=7-9.5 sample. 

\section{Discussion -- Uncovering the epoch of galaxy assembly via pristine gas infall?}\label{sec:disc}

There is now mounting evidence for a potential transition in the physics governing galaxy growth at high redshifts \citep{Heintz23_FMR,Nakajima23,Curti24}, as indicated by the break from the otherwise fundamental-metallicity relation. Previous observations found that this relation was constant out to $z\approx 3$ \citep{Sanders21}, during the majority of cosmic time over the last 12\,Gyr. The break to lower metallicities at high-redshifts has been interpreted as a signpost of chemical `dilution' due to excessive pristine gas inflows \citep{Heintz23_FMR}, as also indicated by the higher fraction of strong Lyman-$\alpha$ \hi\ absorbers in galaxies at these redshifts \citep[e.g.,][]{Heintz24_DLA,Heintz25,Umeda23,DEugenio24,Hainline24b,Witstok24}. 

Indeed, \citet{ZihaoLi25} proposed an analytical model describing the break from the FMR towards lower metallicities and found that dilution from excessive pristine gas accretion was favoured over feedback or outflow effects. \citet{Heintz22} also found observational evidence for the bulk of the intergalactic \hi\ gas being accreted in galaxy halos by $z\sim 2-3$ by comparing the build-up of the \hi\ gas mass in the ISM to the global \hi\ gas mass density inferred from quasar absorbers \citep[see e.g.,][and references therein]{Walter20}. Observations of gas-selected quasar absorption-line galaxies show a sharp break at $z = 2.6\pm 0.2$ in the metallicity evolution with redshift of the sample \citep{Møller13}, argued to signal the transition in the mode of galaxy growth due to the starvation of primordial gas infall. 

While these complementary results seem to support the physical scenario where the offset in the FMR is primarily driven by the physical transition of galaxies being fed new pristine gas via inflows, there are also potential alternative explanations. \citet{Liu25} proposed that the observed high-$z$ FMR can be constructed with a more efficient formation of high-mass ($m_{\rm max}>200\,M_\odot$) stars following a ``top-heavy'' initial mass function (IMF). Interestingly, \cite{Cullen25} find that a top-heavy IMF or other more exotic stellar populations are required to explain the extremely low metallicity of EXCELS-63107. 
Strong outflows, as expected for high sSFRs ($>25\,{\rm Gyr}^{-1}$) like we observe here, may also expel the dust and metals from the galaxies \citep{Ferrara24a, Pallottini24}, which could also explain the overabundance of bright UV galaxies at $z\gtrsim 10$. However, we do not see any prominent signatures of high-mass star formation in our sample, such as nitrogen overabundances, nor any broad emission-line ``wings'' indicative of outflows. 
%(though see e.g. Heintz et al., in prep.)

%These complementary results corroborate the proposed 
%physical interpretation of excessive pristine gas inflow is driving the break from the FMR during the galaxy assembly at cosmic dawn.   
%scenario where the offset in the FMR is primarily driven by the physical transition of galaxies being fed new pristine gas via inflows 

Instead, we argue that the most likely explanation for the offset from the FMR at $z\gtrsim 3$ is representing an excessive pristine gas inflows onto galaxies, which eventually become infall starved and evolve in a near-equilibrium state, mainly re-processing previously acquired gas at later cosmic times. Intriguingly, the peak of cosmic star formation occurs at $z\sim 2$ \citep{Madau14}, approximately 800\,Myr after the transition at $z=2.6$, similar to the typical gas depletion times of star-forming galaxies at early cosmic epochs \citep{Tacconi18,DessaugesZavadsky20,Aravena24}. This suggests that for the galaxies now being found in abundance and characterized in detail with \jwst\ at $z\gtrsim 3$, we are directly witnessing their formation in progress.

% {\bf discuss IMF stuff/paper?}

\section{Conclusion \& future outlook} \label{sec:conc}

In this work, we characterized in detail the rest-frame UV and optical properties of a sample of galaxies at $z=9.3-10.0$, with the main goal to determine their direct metallicities. This is enabled based on the updated reduction and data processing pipeline of \jwst/NIRSpec Prism spectroscopic observations, as part of the 4th version of DJA \citep[see][for previous releases]{Heintz25,DeGraaff24}, extending the wavelength coverage to $5.5\mu$m. Our sample included new observations from major \jwst\ legacy surveys such as CAPERS, RUBIES, UNCOVER, and JADES. Due to the extended wavelength coverage and increasing spectral resolution with wavelength delivered by the NIRSpec Prism configuration, we were able to resolve the auroral [\oiii]\,$\lambda 4363$ line emission and compare to other rest-frame optical line transitions such as the [\oiii]\,$\lambda\lambda 4959,5007$ doublet out to $z=10$. This provided the first direct constraints on the metallicity, ionisation state, and electron temperatures for a sizeable sample of galaxies at $z\approx 10$.

Overall, we found that the sample galaxies were characterized by line ratios indicating strong ISM radiation fields compared to their local galaxy counterparts, with typical electron temperatures $T_e \gtrsim 2\times 10^{4}\,$K. The derived metallicities were on average 10\% of solar, ranging from 12+log(O/H) = 7.1 to 8.3. We found good agreement between existing high-redshift strong-line diagnostics in the literature \citep[e.g.,][]{Sanders24} and our sample, suggesting that these are universally valid at $z=3-10$. We further derived an empirical relation connecting the direct metallicities to the UV brightness, $M_{\rm UV}$, of the targets, valuable to estimate the metallicity for the much more numerous population of galaxies at $z\gtrsim 10$ only covered by \jwst\ photometry. 

We modelled the rest-frame UV to optical SED of each galaxy with {\sc Bagpipes} \citep{Carnall19} to determine their basic physical properties such as the SFR and stellar mass. Jointly with the emission-line analysis, we found evidence for bursty star formation with the SFR on 10\,Myr to 100\,Myr timescale on average being ${\rm SFR_{10}/SFR_{100}} \approx 2$. The SFR-stellar mass main sequence and the mass-metallicity relation of the sample galaxies at $z\approx 10$ were consistent with expectations based on lower-redshift observations, indicating only a mild evolution from $z\approx 6$ to 10. Intriguingly, we found significant evidence for a systematic offset of $-0.41\pm 0.10$\,dex from the fundamental-metallicity relation \citep[e.g.,][]{Mannucci10,Curti20} at $z\approx 10$, supported also by the larger PRIMAL sample \citep{Heintz25}, though based on strong-line calibrations for the metallicities. This indicates that pristine gas infall drives the overall star formation history of the Universe.

To push the current redshift frontier of $T_e$-based metallicity measurements, \jwst's MIRI spectroscopic capabilities could be fully utilised to cover the redshifted auroral and nebular emission lines at $z\approx 10$ \citep[e.g.,][]{Hsiao24b,Zavala25}. The lower sensitivity of MIRI means this would require a large time investment; however it remains our only instrument capable of covering the necessary rest-optical lines above $z\approx 10$. These lines will remain out of reach for next-generation telescope such as the ESO Extremely Large Telescope (ELT), which will instead provide a substantial improvement in spectral resolving power, allowing more detailed characterizations of the rest-frame UV features at $z\gtrsim 10$, which are key to understand the nebular densities and temperatures of the star-forming regions of galaxies at cosmic dawn. 
%or potentially via far-infrared interferometers such as the Atacama Large Millimetre/sub-millimetre Array (ALMA)
 
% {\bf Add future outlook, i.e. how to measure direct metallicities with MIRI, ELT, etc.}

\begin{acknowledgements}
    We express our greatest gratitude to the investigators on the major \jwst\ observing programs, such as RUBIES, CAPERS, UNCOVER, and JADES. The work presented here would not have been possible without their major efforts in designing and obtaining the observational data included in our work here. 

    The Cosmic Dawn Center (DAWN) is funded by the Danish National Research Foundation under grant DNRF140.
    The data products presented herein were retrieved from the DAWN \jwst\ Archive (DJA). DJA is an initiative of the Cosmic Dawn Center, which is funded by the Danish National Research Foundation under grant DNRF140.
    This work has received funding from the Swiss State Secretariat for Education, Research and Innovation (SERI) under contract number MB22.00072, as well as from the Swiss National Science Foundation (SNSF) through project grant 200020\_207349.

%%% JWST ACKNOWLEDGEMENTS %%%
This work is based in part on observations made with the NASA/ESA/CSA James Webb Space Telescope. The data were obtained from the Mikulski Archive for Space Telescopes (MAST) at the Space Telescope Science Institute, which is operated by the Association of Universities for Research in Astronomy, Inc., under NASA contract NAS 5-03127 for \jwst. 
\end{acknowledgements}

\bibliography{ref}{}
\bibliographystyle{aasjournal}

\label{LastPage}
\end{document}